\documentclass[reprint,amsmath,amssymb,aps,pra,superscriptaddress]{revtex4-1}
\usepackage{gensymb,booktabs}
\usepackage{tikz}
\usepackage{graphicx}
\usepackage{url}

\usepackage{breakurl}
\usepackage[unicode=true,breaklinks]{hyperref}
\usepackage{longtable}
\usepackage{graphicx}
\usepackage{dcolumn}
\usepackage{bm}

\date{\today}
\newcommand{\QSERG}{Quantum Systems Engineering Research Group, Loughborough University, Loughborough, Leicestershire LE11 3TU, United Kingdom}
\newcommand{\LBORO}{Department of Physics, Loughborough University}

\begin{document}

\title{The challenges for Systems Engineers of non-classical quantum technologies}

\author{Michael J. de C. Henshaw}
\email{M.J.d.Henshaw@lboro.ac.uk}
\affiliation{\QSERG}
\affiliation{The Wolfson School of Mechanical, Electrical and Manufacturing Engineering, Loughborough University}

\author{Mark J. Everitt} 
\email{M.J.Everitt@lboro.ac.uk}
\affiliation{\QSERG}
\affiliation{\LBORO}

\author{Vincent M. Dwyer}
\email{v.m.dwyer@lboro.ac.uk}
\affiliation{\QSERG}
\affiliation{The Wolfson School of Mechanical, Electrical and Manufacturing Engineering, Loughborough University}

\author{Jack Lemon}
\affiliation{\QSERG}
\affiliation{\LBORO}

\author{Susannah C. Jones}
\affiliation{Defence, Security and Technology Laboratory, Porton Down, UK}

\begin{abstract} Non-classical quantum technologies that rely on
manipulation of quantum states and exploitation of quantum superposition
and entanglement are approaching a level of maturity sufficient to
contemplate commercialization as the basis of practical devices for
sensing, communications, navigation and other applications in the
relatively near-term. However, realization of such technologies is
dependent upon the development of appropriate Quantum Systems
Engineering (QSE) approaches. It is clear that whilst traditional
systems engineering will support much of the integration need, there are
aspects associated with system of interest definition, system modelling,
and system verification where substantial advances in the systems
engineering approach are required. This paper lays out in detail the
challenges associated with Quantum Enabled Systems and Technologies
(QEST) and analyses the adequacy of systems engineering processes and
tools, as defined by the Systems and Software Engineering lifecycle
standard (ISO/IEC/IEEE 15288), to meet these challenges. The conclusions
of this paper provide an outline agenda for systems research in order to
engineer QEST.
\end{abstract}

\maketitle


\section{Introduction}\label{introduction}

Non-classically explainable aspects of quantum physics have been
fundamental to the development of many modern day devices:
semiconductors, the transistor, the laser, etc., underpin the development
of the information (or digital) age in which we live. The application of
quantum phenomena in almost all these technologies has been to improve
upon classically familiar concepts such as Boolean algebra and the
electronic switch (i.e. the transistor). Now however, we are at the dawn
of a new era where state-of-that-art technology promises the ability to
leverage quantum effects for the creation of new capabilities.
Manipulation of the quantum state and specific exploitation of quantum
superposition and entanglement may lead to capabilities such as sensing
through walls, improved navigation (e.g. satellites), GPS-free
navigation, more secure communication, and even quantum computing. These
new devices will involve the creation and manipulation of a delicate
macroscopic quantum state that is likely to require systems designers to
have some understanding of the underlying physics as well as their
engineering discipline, to harness the technologies' intended
capabilities. This new quantum revolution~\footnote{It is appreciated
  that revolutions are usually defined retrospectively and that
  predicting them tends to be an unreasonably speculative endeavor.} has
been well summarized by Pritchard and Till~\cite{1,2}; they
originally used the term Quantum 2.0 to distinguish between mainly
20\textsuperscript{th} Century technologies and those emerging from the
most recent quantum physics research. This term can lead to ambiguities
as usage is not standardized and so this paper will use the expression
``Quantum Enhanced Systems \& Technology(ies)'' (QEST) to refer to
systems that rely specifically on, or make use of, manipulation of
quantum state, as described above. The first commercial devices
exploiting these quantum properties are anticipated within the next few
years; they are likely to be devices that interact with the environment
in communication, sensing and imaging applications. QEST is, therefore,
of interest to engineers who must turn science into workable, safe, and
commercially viable products and systems. But implementation will only
be possible if certain new engineering problems, unique to systems
reliant on quantum coherence, can be overcome. These include new types of
interfaces and new insights into the concept of a system's boundary, the
critical significance of interactions between quantum systems and the
environment in which they must operate, the current lack of suitable
engineering models, and the lack of verification (test and evaluation)
methods for such systems. It is clear that new engineering approaches
will be needed in order to realize the potential of QEST.
 
Translating the results achieved within physics laboratories into
manufacturable, safe, and commercially viable devices and systems will
be a substantial engineering challenge. Successful implementations will
only be possible if certain new engineering problems, unique to systems
that are reliant on upon manipulating macroscopic quantum states, are
addressed. Systems Engineers (whose role is to integrate the
contributions from many other disciplines), in particular, must address
the challenge of creating systems that exploit new effects not
previously encountered and obeying new, but as yet, undefined design
rules~\cite{3}. In this paper, we shall introduce the challenges for
Quantum Systems Engineering (QSE), explore the areas in which
development of systems engineering is required and set forth some
possible approaches that will enable the development of QEST. In doing
so, we are cognizant of the distinction that must be drawn between the
practices of early adopters and the later established processes for
maturing and matured technologies. As Christensen has established,
``disruptive technologies typically are first commercialized in emerging
or insignificant markets''~\cite{4} and it is rarely in the interests of
established organizations to invest aggressively in disruptive
technologies. Our main purpose is to look forward to longer-term
development of QSE and recognize that a ``lighter touch'' of systems
engineering will be appropriate for early adopters. Nevertheless, the
systems thinking approach will also be essential for early adopters in
the task of transitioning from laboratory to real-world application.

Most systems organizations use the ISO/IEC/IEEE 15288 standard,
``Systems and Software Engineering -- System Life Cycle Processes''
\cite{5} to define the processes used in the development and support of
complex systems. The purpose of the standard is to support projects and
organizations in the acquisition and supply of systems. As such, it
provides a comprehensive overview of cradle-to-grave systems management,
and so this standard is used as an analysis framework, through which we
identify the process areas in which QEST quantum phenomena must be
considered, and the types of methods and tools that will be required to
support the engineers charged with developing the systems. The use of
the standard as an analysis framework leads to the identification of
gaps in knowledge and the concomitant research needs for systems
engineering.

It is always the case that the lifecycle must be tailored to the
particular project at hand; Annex A of the standard \cite{5} discusses
this aspect. Tailoring takes account of both organizational needs and
project-specific features; it results in the adoption of particular
processes and tools for, and intensity of, systems engineering.

To begin, we must agree a definition of what we mean by a system and the
important properties with which the systems engineer deals. \emph{``A
system is a combination of interacting elements organized to achieve one
or more stated purposes''} \cite{5}. Thus, a system has elements (which
may refer to parts, components or sub-systems); relationships between
those elements; and a purpose. The relationships can be thought of as
conduits along which material, energy, or information is transported
from one element to another \cite{6}, and these must be organized such
that an emergent behavior will result (i.e. one that cannot be achieved
by any of the elements acting in isolation, but only as a result of
interactions between elements). In the case of engineered systems, we
speak of them having an ``operational purpose'' which satisfies the need
of a customer, operator, or of society in general, and a ``functional
purpose'', which is essentially what the systems must actually ``do'' to
meet that need \cite{7}.

In addition to the System of Interest (SOI, see below), there are
so-called enabling systems that are required to support the SOI in its
functions. Examples could include environment, control systems,
maintenance systems, training, etc. \cite{5}; these have a particular
significance for QSE as we shall discuss below.

Systems Engineering is the \emph{``interdisciplinary approach governing
the total technical and managerial effort required to transform a set of
stakeholder needs, expectations, and constraints into a solution and to
support that solution throughout its life''} \cite{5}. Systems Engineers
must, therefore, concern themselves with the interests of stakeholders
and all the contributing disciplines to a system: their job is to plan
and manage integration.

Maturity of technologies are often assessed using Technology Readiness
Levels (TRLs) that range from ``basic principles observed and reported''
at level 1 to ``actual system proven through successful mission
operation'' at level 9 \cite{8}. It is worth noting that integration is a
part of the activity required to achieve each of levels 5,6,7,8, and 9,
and so Systems Engineering can be considered to be the key aspect of
achieving operational readiness for a new technology\footnote{We are
  indebted to Andy Powell of Innovate UK KTN for this consideration.}.
However, it is important to note the findings of Honour \cite{9}, that
investment in systems engineering early in the lifecycle (i.e. low TRL)
is most significant in terms of project success. It is for this reason
that the Quantum Technologies community has an opportunity to truly
benefit from early adoption of Systems Thinking and Engineering.

Allhoff et. al. \cite{10} commented that ``it is sometimes very easy to
get caught up in what is scientifically possible and ignore the
engineering problems that come with it.'' The purpose of this paper is
to signpost the practical engineering considerations for development of
QEST devices. It starts with a description of QEST and then develops the
modern day challenges of defining a system of interest and how QEST may
affect that. A description of the way in which ISO/IEC/IEEE 15288 should
be used and tailored follows as an introduction to a section by section
analysis of the relevant technical processes that QSE must consider.
Through consideration of current QEST application projects (in the
defense domain), we examine specific challenges associated with
development from TRL 3 to TRL 7. Drawing together the threads of these
various analyses, we conclude with recommendations concerning how QSE
should be developed in order to realize the amazing possibilities that
QEST offers to society.

\section{Non-Classical Quantum
Technologies}\label{non-classical-quantum-technologies}

Many technologies of the electronic age, and latterly the information
age, rely on quantum mechanics to explain and optimize their operation.
Developed in the early 20\textsuperscript{th} Century, Quantum Mechanics
was a departure from classical physics that was concerned with
understanding light (photons), matter (atoms) and their interactions.
Thus, the operation of semiconductors, lasers, and a host of familiar
technologies can be described using the mathematics associated with
quantum mechanics. Pritchard and Till \cite{1} have referred to this as
the \emph{first quantum revolution}, arguing that we are on the brink of
a \emph{second quantum revolution} (QEST) in which devices actively
create and manipulate quantum states either to improve upon existing
capabilities or to deliver fundamentally new solutions. New technologies
will be enabled by directly exploiting the subtle quantum effects of
quantum entanglement, superposition, and tunneling. These effects will
potentially enable extraordinary advances in applications such as
precision timing (clocks), communications, sensing, ,computing \cite{11},
and many others.

\subsection{Subtle Quantum Effects
QEST}\label{subtle-quantum-effects-qest}

\emph{Quantum Superposition:} encapsulated in the famous analogy of
Schr\"odinger's cat, the principle of superposition is that an object
exists simultaneously in more than one possible state -- and for cat
states these are macroscopically distinguishable. Such superpositions
can be used to advantage in information processing and metrology, and
coherently manipulated. They persist until we try to measure the state
(at which point superposition ceases\footnote{This is somewhat of an
  oversimplifications but a discussion of some of the more subtle
  aspects of quantum physics are not within the scope of this paper.})
or environmental effects destroy them (indeed it is their sensitivity to
environmental effects that can make quantum objects potentially
excellent sensors). Superposition implies that an object can be in more
than one place, or have more than one value of an internal variable.
Through super-cooling, scientists have achieved quantum states in
visible objects \cite{12}. This effect is pertinent to some types of
sensor and quantum computing.

\emph{Quantum Entanglement:} is a specific form of superposition between
distinct physical systems and concerns the deeper than classical
correlation of the state of the system (such as position, momentum,
polarization and spin) of objects often separated widely in space and/or
time. Thus, measurement of one object will define another, even though
no information has been exchanged\footnote{Again a necessary
  simplification has been made.}. Entanglement has been demonstrated for
objects as large as small diamonds \cite{13}. This effect is pertinent to
communications and quantum computing.

\subsection{Applications}\label{applications}

It is helpful to summarize the likely applications that will be enabled
by QEST. These are applications that are currently the focus of research
and development programmes, but at varying levels of maturity\cite{1}.
Most of these applications represent technology push, in the sense that
QEST may provide an opportunity to improve an existing capability
(faster, more accurate, etc.), but as experience is gained, it is likely
that entirely new capabilities (as yet unconceived) may be created.

Precision timing is significant for many applications, examples include
Global Navigation Satellite Systems (GNSS), financial networks and
trading, communications systems, satellite monitoring and ground-based
navigation (through GNSS), and electrical power grids. However, atomic
clocks that use existing technologies can already achieve accuracies of
$2 \times 10^{-16}$, at which level the effect of variations in
the local gravity vector have an observable effect in accordance with
general relativity. Finding an application, outside the national
standards laboratories, that can exploit an improvement of 2 orders of
magnitude from a new generation of quantum clocks \cite{1} may be
difficult; perhaps counter-intuitively, the value of the new technology
may be that it makes possible a smaller or lower power solution for
applications that have less demanding timekeeping requirements.

In the area of secure communications, Quantum Key Distribution (QKD) is
an important application of entanglement. Point-to-point QKD is already
available, but development of networks is still experimental \cite{14}
and there are also technical challenges with practical devices
associated with range and room temperature for free space operation;
arguably, now is exactly the time to adopt QSE in order to address such
challenges. Furthermore, QKD is still vulnerable to man-on-the-side
attacks if the attacker can achieve a timing advantage \cite{15}.

Many advances are anticipated in the area of sensors, with QEST effects
offering elegant systems solutions for sensing through walls, detection
of hidden materials or deep underground voids, quantum radar to detect
stealth objects \cite{16,16a}, and ghost imaging,No enabling operators to see
round corners. Entanglement may enable biometric sensors that work at
the sub-cellular level, but without causing damage. Quantum technologies
offer the possibility of several orders of magnitude improvements in
sensitivity for the measurement of electric and magnetic fields.

Further (but less mature) areas for development include navigation using
the local gravitational environment (GPS-free!), gravity imaging for
scene analysis, and detection of movement of massive objects. Rotation
sensors of 2 nrad/s/Hz$^{-\frac{1}{2}}$ sensitivity will offer
benefits, including: inertial navigation, roll stabilization, image
stabilization, various vehicle (e.g. UAV) controls, spacecraft guidance
and survey systems.

Quantum computers were proposed around 1980 and work differently to
traditional binary electronic computers. Potentially, they offer
(for some algorithms) an exponential improvement in computational speed over conventional
computers for some specific forms of calculation that are important in
cryptography and search algorithms, and significantly lower power
demands. Predictions vary for when a viable machine for general use will
be available, but IBM have announced their intention to construct
commercial IBM Q systems with \textasciitilde{}50 qubits in the next few
years

\section{Practical implications for Engineering
QEST}\label{practical-implications-for-engineering-qest}

There are a number of engineering challenges associated with development
of the applications to exploit QEST. Firstly, there is the problem of
maintaining fragile quantum states over extended timescales and
distances; this problem is explored below through considerations of
current QEST application research and development in the defense domain,
though the problems are domain-independent.

Secondly, there are problems of scale and scalability. Whilst these
problems are not unique to QEST, they are also discussed, because they
must also be overcome in order to realize practical QEST devices.

\subsection{QEST Integration Implications, based on current development
projects}\label{qest-integration-implications-based-on-current-development-projects}

The appeal of QEST devices as sensors is their sensitivity to
environmental effects which allows the user to infer specific values
with high precision. However, a fundamental issue is the difficulty in
isolating a single environmental effect, whilst mitigating the coupling
effect to all other environmental factors, which may degrade the
functionality of QEST.

\subsubsection{Implications}

As stated above, a ``A system is a combination of interacting elements
organized to achieve one or more stated purposes'' \cite{5}. A quantum
experiment within a laboratory environment can be considered as a
functioning isolated system, as it has operational elements and/or sub
systems working together in order to achieve a purpose. In this case,
the purpose is to demonstrate the proof of concept of a system (i.e. TRL
3 \cite{8}).

Commonly, these elements are a combination of non-classical
technologies, such as lasers and quantum states, driven by classical
devices, such as power supplies, amplifiers, etc., which are typically
not constrained by size, weight or power requirements. To ensure that
the system meets its purpose it is operated by highly trained physicists
within a laboratory: this reduces the chance of interactions between
operating elements and real world environmental effects.

The developmental step from quantum experiment to QEST prototype (TRL 4)
is non-trivial, as the system's purpose has to evolve from proof of
concept of an experiment, to proof that the system has a real world (by
which we mean to imply an environment outside an experimental
laboratory) operational and functional capability. This imposes loose
constraints on the system's size, resulting in a set of operational
elements that integrate together within a confined volume.

Sensing and time keeping QESTs require cold atoms (100\,$\mu$K)
that can be manipulated into a functioning quantum state. This requires
periods of cooling and state preparation, prior to state evolution,
where an atomic state is allowed to couple to an environmental effect.
The sensitivity of the QEST is dependent on the precision to which these
processes are carried out. Stray magnetic fields, generated by driving
operational elements that interfere with these processes pose a
substantial risk to the functionality of QEST. Therefore, care is needed
when integrating classical devices, such as ion pumps (which generate
magnetic fields), with non-classical elements to preserve quantum
states, in order to reduce sympathetic interactions and parasitic
effects\footnote{Sympathetic interactions and parasitic effects are the
  nomenclature terms used to broadly describe the undesired coupling of
  quantum states to unmitigated environmental interactions.}. Within a
laboratory environment Mu-metals are used around areas of quantum state
preparation to diminish this effect. However, this increases the weight
and volume of a system and may inhibit deployment in real-world systems.
Quantum Technology Shielding is an important area of current research
\cite{17}.

Integrated non-classical driving devices within a QEST can induce a host
of additional undesirable features such as resonant vibrational
frequencies and thermal gradients, all of which can reduce the QEST
functionality and even result in permanent damage to the system, if left
unmitigated.

The next major step in the development of a QEST device is
proof-of-principle that the system fulfils its functional and
operational purpose within its intended environment (TRL 5). With
regards to QEST, the original device is required to operate on a
platform\footnote{A group of technologies, or a major system, used as
  the base upon which applications are operated} subjected to real world
environments. Thus, the QEST device must be integrated as a sub-system
of a platform system such that desirable emergent behaviors of the
platform with QEST will result.

Many of the same issues overcome in developing the laboratory QEST
prototype are apparent in real world environments following systems
integration, except, it is now the platform and its subsystems that
contribute undesirable features, such as temperature gradients,
electromagnetic interference, vibrations and acoustic noise. There is
also the added complexity that the QEST prototype is now constrained in
size and weight by the platform and installation. Above all, the QEST
device must integrate with the platform without interfering with other
subsystems; for some platforms, this could have safety implications.

Cold atom inertial sensor advancements \cite{18} have highlighted the
potential to overcome current navigation sensor issues; specifically
those related to inertial navigation bias instability and drift, related
to Global Navigation Satellite System (GNSS) \cite{19,20}.
However, the technology is currently hindered by the tradeoff between
interrogation time, coherency of quantum state and dead time \cite{21}.
 
Initial prototype testing of an inertial sensor in a micro gravity
airplane flight has indicated the feasibility of a QEST atom-based
gravimeter \cite{22,23}. However, platform vibration levels 
($10^{-2}$g Hz$^{-1/2}$), variations in acceleration ($0-1.8$\,g), rotation rates ($5^{\circ}$ s$^{-1}$) and
systematic uncertainty from time-varying magnetic fields present
significant challenges. In addition to this, separate investigations
have found that radio communication systems on board a test flight
resulted in interference with the QEST laser stabilization \cite{23}.

Submariners could benefit significantly from QEST navigation devices
\cite{18}, however they are one of the most hostile working environments
known today and immature on-board systems could lead to a catastrophic
loss of life. Consequently, there are numerous limiting factors to
consider when integrating QEST devices into this type of environment.

A few well known examples of the predominant factors that could reduce
QEST functionality are:

\begin{itemize}
\item
  thermal variations within an enclosed vessel,
\item
  size restrictions,
\item
  limitations of Radio Frequency (RF) emissions, and
\item
  weight restrictions imposed in order to achieve the correct buoyancy.
\end{itemize}

However, a case study carried out by the Chartered Institution of
Building Services Engineers (CIBSE) \cite{24}, highlights several new
limiting factors that are relativity unknown;

\begin{itemize}
\item
  variance of air pressure, which can be 25\% above and below
  atmospheric pressure,
\item
  onboard ventilation, which can result in large variations in
  concentration of oxygen (O$_2$), carbon monoxide (CO), carbon dioxide
  (CO$_2$) and hydrogen (H$_2$),
\end{itemize}

To overcome the size and weight limiting factors, current active areas
of research are developing chip scale QEST~\cite{25,26}
and portable laser systems~\cite{26,27,27}. However, the
variance in air pressure and environmental gasses could still pose a
significant risk to the functionality of the technology and requires
further investigation.

From just a few examples it is apparent that for early QEST devices,
stepping out of the laboratory to real world environments (TRL 4 to 5)
is arguably one of the most challenging tasks in the near future and the
early adoption of systems engineering approaches is likely to ease this
transition.

An aspect of systems integration that will enable greater understanding
of the trade-off between sensing superiority and engineering limitations,
and illuminate the overall QEST integration challenge, is to develop an
approach to integration of prototype QEST within existing platform
sub-systems, rather than as deployment of completed platform
sub-systems. Such an approach would pose significant new challenges, but
offers the prospect of high payoff. It is envisaged that this will
become an active area of research within the next few years and could
form the next development step in meeting platform operational time and
reliability requirements [TRL 6 and 7].

\subsection{Scale Implications}\label{scale-implications}

Models for macro-scale engineering must be combined with models of
phenomena that occur at relative scales of the order of
10\textsuperscript{-8} to 10\textsuperscript{-16}. The physics is
necessarily different, but phenomena at vastly different scales must
somehow be accommodated. With many applications, laboratory
demonstration is not proven to be scalable to practical devices and so
new models are required to understand the scalability issues. This has
been discussed by Papadakis from the perspective of systems engineering
micro- and nanoscale devices \cite{28} and focuses on the consideration
of changes in the behavior of continuous variables (material and fluidic
properties) and those that give rise to quantized behavior, but mainly
drawing attention to the need for quantum mechanics to explain certain
phenomena. In terms of the impact of scale on systems engineering,
Sample \cite{29} has drawn attention to the variability in batches of
nanomaterials, having implications for integration and asserting the
need for standardization. She noted, in particular, that higher failure
rates might be anticipated and this certainly has implications for the
reliability aspects of systems engineering. However, Breidenich \cite{30}
has focused on interfaces as being a critical issue for Systems
Engineering of micro and nanotechnologies, because the variation in
scale adds considerably to the complexity of defining interfaces.

So far, although scale is recognized as a major issue for systems
engineering, there are no definitive views on how it should be dealt
with in general. This is an area of systems engineering research with
significant implications for the practical realization of QEST devices
and systems.

\subsection{System of Interest (SOI)}\label{system-of-interest-soi}

In order to design, develop, or update a system, it is essential (for
philosophical, technical and commercial reasons) to define the system
clearly. \cite{6} describe the `Narrow System of Interest' (NSOI) as
those elements of the system directly under the control of the engineer;
the `Wider System of Interest' (WSOI) as those elements of which the
engineer must take account, but which are not directly under the
engineer's control; the `Environment' that provides context for the SOI;
and the `Wider Environment' that has no influence at all on the system
(see Figure 1). They also discuss the `Meta System' which could be
considered as the organizational and technical system used to build or
develop the NSOI. For simple systems, the NSOI can be defined in a
straightforward manner as the elements that the system contains: the
analysis for this is referred to as `structured system identification'.
For complex systems, `behavioral systems identification' is a more
appropriate approach, in which the interactions of interest are
identified. The engineer must clearly classify the elements of a system
as within NSOI, WSOI, environment, or wider environment in order to
understand how the elements should be treated.

In recent times, increases in technology system complexity and
significant increases in connectivity has led to difficulties in
identifying the NSOI. Systems of Systems (SoS) considerations mean that
many different properties must be taken into account \cite{31} and the
dynamic nature of systems (in which connections appear and disappear)
must be represented somehow in the definition of the system \cite{32}.

The subtle quantum effects described in section II.A introduce
non-traditional interactions that generate significant problems in
distinguishing between NSOI and WSOI and may, furthermore, bring
elements of what may have previously been considered as environment into
the SOI. For instance, when measuring the rotation of a platform, one
must also consider the rotation of the earth; also, when measuring
gravity (wider environment), one must account for local
environmental movement. For quantum systems, interactions with the
environment are very significant because they can destroy the quantum
states that allow QEST behaviors within microseconds. Quantum
systems that use entanglement to achieve their purpose will need to
somehow include the entangled objects within the NSOI, but it is not
clear how this should be done. For instance, moving entangled atoms
around ``components'' in an atomic computer challenges the usual notion
of system boundaries and design methodologies.

Furthermore, the coupling between enabling systems and the NSOI may also
lead to ambiguity in definition of the systems boundary. For instance,
the system to laser cool and trap atoms for certain applications
\cite{33} could be regarded as an enabling system and, therefore, in the
WSOI. However, in practice this may require development within the
overall application and thus fall within the NSOI. For experiments in
the laboratory, this is not an issue, but for practical, commercial
devices, this becomes an important consideration for costs and workflow.

The SOI is conceptual and its articulation by a systems engineer
reflects the how they understand the system. For complex systems,
explicit definition of the SOI is the means through which co-workers may
develop a shared understanding. Whether explicitly stated, or simply
implied, the SOI defines what the developer includes in models of the
system, which has a direct bearing on how the system is understood. In
practice, a systems engineering team will be able to reach an agreement
about the SOI for QEST, but there is clearly a need for principles to be
developed to enable a consistent and meaningful approach to definition
of SOI for QEST.

\begin{figure}[!t]
\includegraphics[width=1.0\columnwidth]{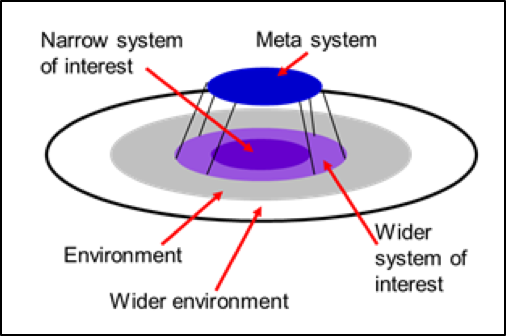}

\caption{Top Hat model of System
of Interest, as conceived by R.L. Flood \cite{34}.}

\end{figure}

\subsection{Challenges}\label{challenges}

To summarize, quantum systems engineering must consider:
\begin{itemize}
\item Integration of nanotechnology and other non-conventional technologies
(such as low temperature or ultra-high vacuum) due to properties of the
materials \cite{29}

\item Multi-scale issues, especially concerning the combination of models
from vastly different scales \cite{28}, which makes partitioning and
interface specification challenging

\item System identification with reference to the challenges of subtle QEST
effects and to the multi-scale issues above

\item Scalability in terms of expanding experimental technologies to
practical devices: it is not a foregone conclusion that effects will
scale in a straightforward manner, or at all

\item Emergence such that decomposition and analysis of a system may not
enable overall behavior to be predicted at all \cite{35}. This is a basic
problem of the complexity of the systems and the scientific uncertainty
of the quantum interactions

\item Interface specification: related to the emergence problem and the
multi-scale issue above, specifying interfaces to include QEST effects
and, perhaps more particularly, understanding where the interfaces occur
between different systems, or different components, will be a significant
challenge

\item Device size, robustness and environment: for many potential
applications the associated enabling systems (e.g. cooling systems) are
of substantial size and the environmental demands are challenging.
Furthermore, precision alignment of components means that robustness in
non-laboratory environments is demanding. This means that integration
into practical environments (e.g. an airplane) will require
technological advances

\item Design approach: Systems Engineering is traditionally thought of a
top-down approach (although frequently bottom-up also occurs); related
to the multi-scale issue identified above is the need to combine a
top-down and bottom-up approach to design \cite{36}. Also, if the systems
engineering community is to identify additional applications for QEST it
will need to do considerable `bottom-up' modelling to be able to
identify and articulate the benefits that could be realized from early
adoption of QEST for each application considered.
\end{itemize}

Although several of the issues above have been considered elsewhere,
there are so far no definitive systems engineering approaches that have
been identified. However, the subtle QEST effects have so far received
little attention from the systems engineering point of view and this
will be the main focus of the analysis below.

Although out of scope for this paper, it would be remiss not to mention
that QEST will rely on the availability of metals such as rubidium, and
the reliability of supply in a congested, contested, and competitive
world may become a critical issue for future exploitation of QEST (see
for example: \cite{37}). Thus one might anticipate that end-of-life and
re-use processes could become a significant part of QSE.

\section{ISO 15288}\label{iso-15288}

The ISO 15288 standard \cite{5} assumes that a typical system life cycle
includes the following stages: concept, development, production,
utilization, support and retirement. Of course, utilization and support
are generally considered to run concurrently. However, the standard
makes no assumptions about the development life cycle structure, which
is chosen by engineers to suit the type of systems development being
addressed. The standard is written mainly from a project perspective and
clusters processes within four categories: agreement processes are
concerned with acquisition; organizational project-enabling processes
are mainly concerned with responsibilities and resources; Technology
management processes are concerned with planning, decisions, risk, and
information management. Technical processes are concerned with the
engineering aspects for system development, sustainment, and disposal;
it is within this category that the special considerations for QEST
occur. In all system development, the life cycle must be tailored to the
particular system of concern: this includes the structure of the life
cycle stages, the processes and decision gates within those stages. In
the sub-sections below, we consider particular processes (identified by
name and clause number from the 2015(E) issue of the standard) for which
QEST considerations should be included. This paper provides guidance to
the systems engineer to where life cycle tailoring should include
specific QEST implications. The development of systems that exploit QEST
is likely to go through several maturity phases within a short space of
time and so it is important to recognize that the systems engineering
methods, processes and algorithms will also need to develop in line with
the maturing levels of quantum integration.

Some familiarity with ISO/IEC/IEEE 15288 will be beneficial to
understanding these sub-sections. It should be assumed that clauses not
listed below have not, at this time, been identified as requiring
special consideration in QSE.

\subsection{Stakeholder needs/requirements definition (clause
6.4.2)}\label{stakeholder-needsrequirements-definition-clause-6.4.2}

That the system exploits QEST effects might be a System Non-Functional
Requirement \cite{38} for a demonstrator device, but it ought not to be
assumed that QEST is the ``answer'' per se. A crucial part of these
processes will be to establish the critical performance requirements and
measures, such that it is possible to decide whether QEST offers
effective benefit over other (more conventional) technologies.
Similarly, the requirements for enabling systems must be established,
although it is possible that these will only emerge after substantial
design work. The acceptability of the enabling systems (size, cost,
capabilities) must be established as early as possible to determine the
viability of using quantum technologies within the application under
consideration. Thus far, there is no specific difference for QSE, but
definition of the down-select criteria in order to prioritize customer
requirements and assess feasibility is a crucial step for quantum,
because many of the factors may currently be unknown or, at least,
uncertain. Until more is known about the practicalities of
implementation, the definition of down-selection must be carefully
constructed to ensure that relevant parameters are measurable and
provide sufficient information for decisions to be made.

To some extent, it could be argued that research in quantum systems
falls into the ``technology push'' class of innovation (i.e. that the
systems provide opportunities for invention of new capabilities),
whereas stakeholder requirements management is predicated on the notion
that stakeholders already know the capabilities they need and wish
suppliers to create appropriate systems to deliver those capabilities.
Currently the type of benefits that quantum systems may provide is
insufficiently defined. Thus, quantum systems may not be considered if
existing techniques will satisfy the need, particularly given that the
technical risks of using them may be high, although this may be
different for early adopters, as noted in section I. Research and the
anticipated effort to exploit quantum systems over the next few years
will provide the knowledge needed to define requirements appropriately
for quantum systems to be considered in the solution space. Furthermore,
various demonstration programs being funded around the world will
develop the skills, experience, industrial capabilities and embryonic
supply chains that will provide the essential infrastructure necessary
for quantum systems to be considered in the solution space.

An example of how quantum technology can provide an unexpected benefit
emerged during the work of one of the authors with the geophysics
industry. Initial discussions established that the sensitivity of the
quantum gravity sensors that are becoming available does not provide
sufficient advantage for users to justify the cost and risk of moving
from the conventional systems that they currently use. However, when the
discussion explored the tasks that the geophysics industry was
undertaking, rather than focusing upon the sensitivity of the sensor, it
became apparent that the rapid stabilization and ruggedness of the
quantum sensors would provide a large benefit to the users. This was a
benefit that could be quantified in terms of the time and cost to
complete each task and thus it made QEST systems an attractive
proposition for the industry.

In summary, the part of the stakeholder requirements processes need
specific consideration is the tailoring of the down select criteria to
ensure that appropriate distinctions between QEST and non-QEST solutions
can be made, and thus quantum benefits are not overlooked.

\subsection{Systems requirements definition process (clause
6.4.3)}\label{systems-requirements-definition-process-clause-6.4.3}

This process transforms the stakeholder requirements into a set of
(technical) system requirements that a proposed solution must meet. The
resulting systems requirements cover functional, non-functional,
performance, process, and interface aspects, and include design
constraints. However, `they should not imply a specific implementation'
\cite{5} and herein lies a dilemma: the use of QEST effects implies a set
of constraints not present with other implementations (e.g. management
of entanglement, temperature limitations for some technologies, size
implications for the whole system). There will also be intrinsic
requirements concerned with interference or non-interference due to QEST
effects. Enabling systems will also be defined at this stage.

Techniques such as Quality Function Deployment (QFD) may be employed at
this stage: this comprises a set of linked matrices that begin with the
stakeholder wants and needs and potential technical responses (Matrix
1). Through a system of scoring appropriateness of the technical
response to the stakeholder need and the correlation of various elements
of the technical responses, the system requirements are gradually
developed. New matrices are added starting from the outcome of the first
to add increasing detail to the technical response. The rules around the
successive matrices vary from one author to another or, more
importantly, one project to another. Cohen \cite{39} has provided a
variety of possible definitions of the matrix of matrices (i.e. linked
set) that address different challenges. \cite{38} has considered four QFD
phases as shown in Figure 2; if one considers the phases to be
iterative, rather than strictly flow down (which may be practically true
in many projects), and one retains several potential technical responses
at the earlier stages, then one may explore the overall solution
requirements, bearing in mind, for instance, that manufacturing
requirements may demand changes in the earlier stages for practical
reasons. The difficulty with this approach for QEST is that a) the
appropriate models for analysis of technical requirements are immature,
because of multi-scale problems and b) a corollary of this is that
emergence is not properly predicted because of the decomposition process
itself. Although QFD is not the only technique that may be used, it is a
popular and effective tool that illustrates some of the challenges
associated with the systems requirements definition process.

The challenges may be partially resolved by development of a top-down
meets bottom-up approach, but the main lack is adequate models of the
potential systems to provide reliable correlation between the elements
of the technical responses.

An outcome from an NPL workshop on quantum timing, sensing, and
navigation \cite{40} was that a family of models is required. A system
model to provide both industry and academia with a basic tool to predict
the system performance that, in principle, could be achieved with the
new generation of sensors in different system configurations. A detailed
system model or simulation that is based on the engineering reality and
can therefore provide realistic estimates of likely performance for a
first generation system employing quantum sensors. This model or
simulation is an essential prerequisite for significant investment in
developing and demonstrating a real system.

\begin{figure}[!t]
\includegraphics[width=1.0\columnwidth]{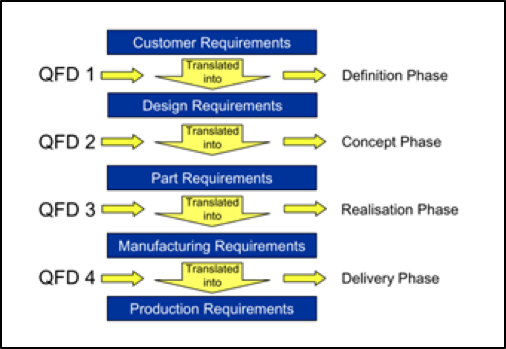}

\caption{Four phases of QFD and
described by Burge \cite{38}.}

\end{figure}

Definition of design constraints due to QEST effects may also prove
problematic due to model limitations.

In summary, tailoring of the systems requirements definition process
should adopt a top-down/bottom-up hybrid approach. A major issue for
these processes is the adequacy of the models used to establish the
relationships and correlations between different technical requirements
and this will only be resolved through the development of better
multi-scale models that properly account for QEST effects (see section
V).

\subsection{Architecture definition process (clause
6.4.4)}\label{architecture-definition-process-clause-6.4.4}

Because a key outcome of architecture definition is the traceability of
the architectural elements to stakeholder and system requirements, the
competence of the architecture definition process clearly has a
dependency on resolution of the issues identified in the previous two
sub-sections. Basically, architecture can be thought of as the
`organization of resources' \cite{41}, but the incorporation of QEST
effects within those resources may require new approaches to the definition
of the properties, characteristics, functions and constraints allocated
to the architectural elements. This will require Quantum Systems
Architects to reach a consensus and establish standard representations
for QEST entities. These will also be applicable to the design
definition processes discussed in the next section. The degree to which
new types of entity are introduced at the architecture level is
dependent, of course, on the level of abstraction that is needed, and
this is not yet clear. The functional architecture is unlikely to be
affected by the introduction of QEST, although specific functions within
that architecture may be new.

There are, however, some architecting principles that should be borne in
mind. Quantum technologies are expected to advance rapidly during the
next few years, as such, one can anticipate the need to upgrade and
improve systems frequently until reliable and stable patterns emerge for
specific applications. Architecting is a key element of system life
cycle planning, and the principle of modular architecting will be very
important. These processes include the assessment of candidate
architectures, which is an immature area within systems engineering
\cite{42}, although developing. The assessment should prioritize the
competence of the modularity for enabling frequent insertion of new
technology. To support the Quantum community, the adoption of open
architectures (which, simply expressed, means that they are published in
sufficient detail for developers to reliably interface their systems to
other systems) should be encouraged \cite{42}.

Definition of system of interest and interfaces will be problematic for
system architecting (see section III.C). Interfaces to the enabling
systems and the specification of enabling systems will require
particular attention. It is usual for interfaces to be specified in an
Interface Control Document (ICD) \cite{43}, although it could be a
reference model rather than a document. A standard template is usually
employed for the ICD in a particular domain and so an ICD template that
includes quantum-related aspects of the interface is required.

The physical architecture will need to take into account the precision
and interaction requirements of the Quantum technologies. As before, the
key issue with this is the adequacy of the multi-scale models for
defining these interactions.

In summary, community effort is needed to define standard representation
of QEST architectural elements. ICD templates for systems employing
quantum technologies are also required. Tailoring will require
particular attention to assessment of candidate architectures that
exhibit good modular and open characteristics. The physical architecture
must take quantum interactions into account to avoid undesirable
emergent behaviors.

\subsection{Design definition process (clause
6.4.5)}\label{design-definition-process-clause-6.4.5}

Following from the architecture definition, the design definition
processes provide sufficiently detailed information about the systems as
a whole and the individual system elements to enable implementation
\cite{5}. The matters and issues identified for that architecture
definition process are also included in this process. However, an
important aspect of this process is the selection of the technologies
required for each system element. Essentially, the inclusion of QEST
increases the trade space available to the engineer in making technology
selection, but herein lies a challenge. There are a large number of
trade space tools available, but as yet no commonly agreed set of models
and processes \cite{44}; often the application of such tools and their
parametrization is proprietary, as this can be a significant component
of a company's commercial advantage. The development of suitable models
and their incorporation into trade space decision support frameworks
should be an area of intensive research effort in the future.
Furthermore, the effectiveness of trade space analysis, even assuming
the models are adequate, is highly dependent on choice and structuring
of the attributes against which the trade is conducted. So far, little
knowledge exists on how to structure the trade for inclusion of QEST
technologies and, once again, the vastly different scales that are
involved in practical devices means that multi-scale analysis will be a
complication. Trade studies take place at various stages of the
development life cycle, at different levels of detail. An important
consideration for tailoring will be the appropriate level of quantum
modelling that must be included at any particular stage of the
development life cycle.

Standards are emerging for quantum technologies, particularly in the
area of security, e.g. quantum cryptography \cite{45} and quantum key
distribution \cite{46} from ETSI\footnote{European Telecommunications
  Standards Institute}; design must necessarily take account of these
standards, but there is a need for additional standards as these
technologies begin to be deployed in practical devices.

It is important to recognize that the individual system elements are
likely to be developed by several or many different organizations. The
processes must account for appropriate supply chain management,
particularly during trade-off studies to ensure that designs are optimal
and can be implemented as designed once production begins.

In summary, tailoring of the design definition process draws on the
recommendations for architectural definition, together with use of
enhanced trade-off tools that specifically take quantum effects into
account. Standards are emerging that will assist the design process
through constraints and interface specification. Trade-off studies
involving the supply chain are already commonplace, but appropriate
education of supply chain members in quantum technologies will be
required.

\subsection{System analysis process (clause
6.4.6)}\label{system-analysis-process-clause-6.4.6}

At a stakeholder meeting for Quantum Systems Engineering held in
2015\cite{47}, attendees were asked: how much quantum mechanics should
engineers need to know to implement quantum technologies?
Unsurprisingly, perhaps, industry representatives responded: ``as little
as possible.'' The question remains unresolved, but it is the demands of
the system analysis process that will determine the level of quantum
mechanics knowledge that systems engineers require. More generally, the
development of QSE techniques should be predicated on finding ways for a
wide range of technical staff, with different specialist knowledge, to
collaborate, and avoiding the need for most to have an in-depth
understanding of QEST. Careful partitioning and the development of
parametric models for the individual components are the usual first
steps at the system level -- some of the component level techniques --
e.g. QFD -- can provide a useful framework for enabling collaboration
across the different skills and disciplines that are essential in
developing and manufacturing a practicable QEST component. Clearly,
quantum physics knowledge is required for the process of validating
assumptions and results.

\subsection{Implementation process (clause
6.4.7)}\label{implementation-process-clause-6.4.7}

Conventionally, implementation refers to the development of system
elements (ready for integration) that meet the requirements,
architecture and design. It is not uncommon for implementation to
introduce new constraints that must then be reflected back into
requirements, etc. Crucial design features that enable delicate quantum
effects will most likely be not subject to compromise, and so the
process must include rules and tests to ensure that unwanted changes are
not introduced. An example might be that even seemingly insignificant
changes to a component could change the thermal paths for the whole
system, destroying the quantum effects or leading to component failure.
One can expect that, as in the case of integrated circuit design, which
is driven by the ITRS\footnote{International Technology Roadmap for
  Semi-conductors} roadmap \cite{48}, (addressed at major international
conferences of IRPS, ESREF and ICMAT and is supported by bespoke tools
in the workflow), reliability will become a major area of research in
quantum technologies. The principles derived from this research will
inform the implementation strategies developed during the implementation
process.

In summary, tailoring of the implementation process will require
principles for reliability engineering to be incorporated. Testing
implementation changes for their potential effect on the whole system
will require modelling in advance of agreement to changes.

\subsection{Integration process (clause
6.4.8)}\label{integration-process-clause-6.4.8}

Assuming the ICD (described in section IV.C) includes the information
relevant to quantum systems, then the integration process should not be
dissimilar to that for regular systems. Similarly to the
implementation stage (IV.F), any constraints introduced during
integration and the process must ensure that these do not affect quantum
effects or system robustness. Integration must take place in an
environment representative of the intended deployment environment to
ensure that the system will interact appropriately with it.

A major challenge to the introduction of quantum technologies is the
size, complexity, and (lack of) robustness of the systems and enabling
systems. This is, however, a challenge for Quantum Systems research,
rather than a change to process.

The application of an incremental development model \cite{7} could be
problematic, because of unpredicted emergent quantum behaviors during
integration of new capabilities, but this model generally assumes that
all the requirements are well-known in advance, so it is unlikely that
this would be a likely choice of development model, especially for early
adopters.

\subsection{Verification process (clause
6.4.9)}\label{verification-process-clause-6.4.9}

Verification establishes that the system, or system element, fulfils its
requirements and is built as it was designed. This is established
through test. The difficulty for testing system elements is that the
fragile states of QEST mean that measurement can be counterproductive.
To carry out verification of elements, there is a need to design for
test, but the tests themselves may be impossible to carry out in
practice. To state it bluntly, because of the observer effect, there is
no point in measuring a quantum property if that measurement destroys
the property. Considering this problem at a more remote level of
measurement, one could check consistency of input and output through
some form of controlled experiment, but even then, it would not be
possible to verify that the system had followed the algorithms that were
intended (i.e. it may be possible to measure end states, but not
intermediate states). It should be noted, though, that there is a large
body of work on certified quantum non-demolition measurement that
addresses this problem from the physics perspective (see for example:
\cite{49}).

The increase in the number of possible system states is also
problematic, because testing must be affordable and time limited.

The interconnectedness and increasing levels of autonomy in modern
systems are challenging conventional approaches to verification. To some
extent, QEST simply adds to the difficult task of verifying complex and,
often, non-deterministic systems. The tailoring of the verification
process requires entirely new approaches to verification; some have
argued that a new verification paradigm is required and work is underway
to develop new tests for quantum systems to support verification e.g.
the QuProCS project sponsored by the European Commission\footnote{\url{http://cordis.europa.eu/project/rcn/193797_en.html}}.

Verification takes place at each step of the systems development life
cycle. When requirements are generated (see sections IVA/B) then a test
that will determine whether the requirement has been met should be
determined. This then becomes part of the test program of the overall
verification plan. For many steps there will be no need for a different
approach to verification, but there will be steps associated with system
element that employ QEST and overall verification of the system, where
testing will be hard to define.

In summary, tailoring the verification process could be the most
intractable parts of the overall life cycle tailoring activity. This is
in part due to system complexity and lack of deterministic behavior of
highly connected systems, and in part due to potentially untestable
states of the quantum system. Together with other systems communities,
there could be a need for an entirely new verification paradigm, e.g.
\cite{50}, though there is not yet agreement on what that paradigm might
be.

\subsection{Validation process (clause
6.4.11)}\label{validation-process-clause-6.4.11}

Perhaps because V\&V is alliterative, validation is often considered to
be part of the same process as verification, but in fact these are
different processes that serve different purposes within the overall
life cycle. Whereas verification is concerned with the correctness of
the system in terms of compliance to design and ensuring that results of
tests meet the expected results, based on system requirements,
validation refers to checking that the system, once built, meets the
stakeholder requirements (i.e. the business or mission objective). If
the stakeholder requirements include a specific quantum requirement,
then this could prove complicated to test, but it is unlikely that the
requirements at the business or operational level would do so; thus,
tailoring of the validation process is unlikely to include specific
quantum aspects. For early adopters, the customer requirement is likely
to evolve so that validation is complicated by not knowing what is
actually required.

\subsection{Maintenance process (clause
6.4.13)}\label{maintenance-process-clause-6.4.13}

The maintenance process will likely need to include monitoring of the
quantum systems so that fault diagnosis is informed and, certainly in
the case of early applications, knowledge of the system behavior
relative to maintenance requirements can be established.

\subsection{Disposal processes (clause
6.4.14)}\label{disposal-processes-clause-6.4.14}

The systems are likely to contain a number of high value components, but
it is unlikely there will be additional considerations for disposal that
are specific to quantum.

\section{Models}\label{models}

The design of any advanced technological system requires a hierarchical
family of models. The hierarchy is usually based on levels of
abstraction, working from a high level model with little detail down to
models which may contain thousands or millions of objects. In systems
engineering the hierarchy of models can mimic, but is not necessarily
equivalent to, the increasing detail associated with requirements
analysis and specification. Scientists and engineers use models to
represent aspects of the world for a variety of purposes \cite{51}, and
in many engineering problems, the family of models will also include
individual, discipline-specific models that may or may not be coupled,
and which serve different purposes within the overall design activity.

For QSE the family of models will need to include, at least, a basic
system model to provide both industry and academia with the means of
predicting the system performance and behavior of a new generation
device in different system configurations, and a detailed system model,
based upon engineering reality, capable of providing realistic estimates
of likely performance.

Focusing specifically on quantum devices, simulation of these can be
split into several potentially overlapping, somewhat subjective
categories:

\textbf{i.} Physically realistic models of the device and its quantum state
(possibly including environmental effects, measurement, feedback and
control). These may be quite sophisticated and realistic but, in the
absence of a quantum computer to solve them, limited to relatively
simple systems. Such models will usually be semi-analytic or entirely
numerical. Indeed, to the best of our knowledge the Berkley Lab
supercomputer boasts the world record of modelling 45 qubits (with a
maximum of 49 simulate-able). As a guide, if Moore's law continues to
hold, an extra qubit would be classically simulate-able every two years.

\textbf{ii.} There is a range of models that try to approximate the state of a
subset of quantum systems including many-body statistical methods and
approximate methods treating quantum systems coupled to an environment
of other quantum systems (open systems methods). While these models have
a quantum nature to them they do have limitations; for example standard
ways of modelling open quantum system require a number of approximations
to be made to derive a dynamical model for the evolution of the systems
state (known of as a master equation). Here a number of choices need to
be made that depend on motivation and which often carry subtle
implications. For example, demanding conservation of probability (i.e.
insisting on Lindblad form \cite{52}) may lead to the introduction of
non-physical additions to the model\cite{Duffus1}\cite{Duffus2}. This approach is often acceptable,
and sometimes the only one available, when seeking to understand the
physics of an open system. However, the value contributed by these
approximation methods has yet to be established.

\textbf{iii.} Quantum models that try to predict behavior of parameters,
characteristics or outputs \cite{53,54}.

\textbf{iv.} Semi-classical and classical models that are similar in ambition to
‎iii, but only in situations where the devices are not really behaving
quantum mechanically.

\textbf{v.} Some exploration of hybrid models has been made that couple, for
example, classical and quantum dynamics \cite{55}. Such approaches have
been used with limited success to model some hybrid quantum-classical
composite systems. There are a number of ontological difficulties with
these models which perhaps explains their limited use in the physics
community and their predictive nature has yet to be established within
the scope of a device engineering perspective.

One of the main motivations for pursuing quantum computing and
simulation is to make possible computing tasks that are not merely
beyond the reach of existing hardware but also beyond the reach of any
foreseeable classical computing solution. Somewhat ironically this level
of computational power is required to carry out quantum device design
using some of the modelling techniques outlined above.

Clearly, there are open questions regarding the most suitable modelling
choices for quantum effects, but these models must then be brought
together with engineering models for system design. These may be
architectural in nature, but the coupling with physics models is by no
means defined. Capabilities analogous to the semiconductor community
will be needed. In CMOS, quantum physics is used to characterize new
transistor designs with models operating at atomic scales, but once a
classical switch has been characterized, hierarchical modelling can then
be used to enable full chip design \cite{56}. Quantum systems are
different; simulating one of any appreciable size can only be
effectively done on a quantum computer (testament to this is the
creativity that the quantum chemistry community has employed in
developing useful models of molecular systems). It seems clear that the
modelling solutions effectively deployed in areas such as CMOS are
unlikely to become part of a systems engineering solution without some
significant additional methods being developed. In CMOS systems, there
is little physicality in the higher level models within the system -
these are concerned with timing and system integration difficulties.
These models are not phenomenological in the way that a physicist is
used to thinking: as the physical implementation is below the level of
abstraction, the models are more process-ological (to coin a phrase).

Giere has stated that models are used for a variety of purposes \cite{51}
and there is a danger, in developing engineering models that rely on
those currently used by physicists, that the motivation (and purpose) in
each discipline is different. Physicists seek to understand nature and
test this understanding in laboratory experiments that may focus on
precision within only a few experiments, whereas engineers are
interested in developing a product that will need to be manufactured,
maintained and operates with a cradle-to-the-grave mindset. As such the
notions of model validity, correctness and fitness for purpose are
different.

Simulation of quantum technologies is, unsurprisingly, one of the
principal challenges of developing the hierarchical modeling systems
engineering capabilities. It is needed to realize the kind of
engineering design and system verification strategies effective in more
established disciplines. This hierarchical notion of modelling has yet
to be realized for quantum systems where it is not clear how quantum
effects such as entanglement can be accounted for at higher levels of
abstraction. As such this represents an area in need of significant
research and development.

To summarize, there is an absence of modelling techniques for QEST
design; models of quantum devices that currently exist are generally not
established for any but the simplest configurations. The linkage between
the physics models and engineering models that may be used to build
design tools is also an area requiring significant research. The lack of
suitable models is probably one of the most significant difficulties
facing QSE, particularly as this impacts design, verification, and
(possibly) validation parts of the system development life cycle.

\section{Discussion }\label{discussion}

\subsection{Identification of key knowledge
gaps}\label{identification-of-key-knowledge-gaps}

Systems engineering combines systemic thinking (consideration of the
system as a whole and its interaction with the environment) with
systematic viewpoint (concerned with the parts of the system and how
they interact) \cite{57}. This paper is predicated on the assertion that
a system engineering approach will be needed to transform experimental
QEST into viable devices, where viable implies that they are
sufficiently robust and reliable to be used for important endeavors and
that the producers of such devices have a realistic commercial
proposition. We have intentionally avoided making an assertion about the
intensity of the systems engineering approach that should be used, but
have recognized that this will be somewhat different for early adopters
(in the next few years) compared to the larger scale enterprises we
anticipate once the QEST begin to mature as commercial systems.

ISO/IEC/IEEE 15288 \cite{5} is the standard recognized by the systems
engineering community as defining the process types for managing the
whole lifecycle of a system. Using this as an analysis framework, we
have essentially asked the questions:

\begin{itemize}
\item
  Does the capability currently exist to perform each technical process
  stage if the system under consideration contains QEST? If not, what
  are the capabilities that must be developed?
\end{itemize}

The approach taken to address the questions has been to assess current
methods, as reported in the literature, and in the experience of the
authors, to determine the extent to which it is likely to give a
satisfactory outcome at each stage. Although this might be considered
somewhat subjective, the purpose is to identify potential weaknesses
that would arise using current methods within a systems engineering
framework, rather than to rigorously determine failures in a
quantitative fashion. Indeed, the potential weaknesses identified
indicate that such a rigorous analysis is not yet possible. The outcome
of this analysis is an agenda for research and development that will
lead to the principles and methods for quantum systems engineering.

Where knowledge gaps have emerged, these are mostly not specific to the
overall maturity stage of QEST.

The detailed discussion of the challenges associated with QEST at each
process stage is summarized in Table 1.

It is clear that the most significant current deficiency is a lack of
appropriate models through which QEST systems can be designed. However,
the difficulty in defining a suitable SOI is a contributor to this
problem, because without a consistent and reliable approach to
definition of SOI, it is not possible to define a model, even if
suitable modelling techniques are available. These two deficiencies,
which are undoubtedly the most significant and most complex that must be
addressed, are of a fundamental nature and must be investigated through
collaboration between scientists and systems engineers.

Other deficiencies (e.g. new architectural representations, integration
strategies, etc.) fall more naturally into the realm of pragmatic
engineering development; these problems can be addressed starting from
current practice. However, it is important to note that the pragmatic
approach may only be possible once progress has been made on the
fundamental aspects.

\begin{table}
\caption{ Summary of QEST
Challenges for Technical Process Stages of ISO/IEC/IEEE 15288}
\begin{center}
    \begin{tabular}{p{0.3\columnwidth} p{0.7\columnwidth}}
    \toprule
\textbf{Process stage} 
    & 
\textbf{Key issues for QEST}
    \\
\midrule 
A.\hspace{4pt}Stakeholder requirements        &  Lack of suitable models for down selection of  
                                      technological approach;
                                      Lack of criteria for down selection to 
                                      include QEST options;
                                      At current maturity of QEST, it is unlikely that                     
                                      stakeholder requirements will be sufficiently 
                                      detailed\\
B.\hspace{4pt}Systems requirements definition &  Lack of suitable models, leading to inability to 
                                      adequately predict emergent behaviors   \\
C.\hspace{4pt}Architecture definition         &  New architectural entities and representations may be 
                                      required for QEST (this is currently tractable); Evolvable
                                      architectures are required (this is currently tractable); Interface
                                      definition may currently be inadequate   \\
D.\hspace{4pt}Design definition               &  Lack of models for use in the trade space; Not currently clear 
                                      how to construct trade space to include QEST\\
E.\hspace{4pt}System analysis                 &  Need to ensure partitioning of design enables specialists can
                                      collaborate without the need for in-depth QEST knowledge
                                      (should be currently tractable)\\
F.\hspace{4pt}Implementation                  &  Principles for reliability engineering must be
                                      included \\
G. Integration                     &  Integration strategies must accommodate quantum effects
                                      to avoid robustness issues\\
H. Verification                    &  Cannot measure intermediate states; Test plan cannot include all    
                                      possible states (too many), implies need for new verification
                                      paradigm\\
I. Validation                      &  Dependent on well-defined stakeholder requirements: this
                                      may not be possible for early adopter QEST devices\\
J. Maintenance                     &  Suitable monitoring strategies need to be defined, but
                                      not yet clear at what level\\
K. Disposal                        &  No issues identified\\
\bottomrule
    \end{tabular}
\end{center}
\end{table}

Of the process stages, it seems that verification is the most
problematic: the measurement of intermediate states is prevented by
physics. Thus the impact of this must be assessed and alternative
approaches to verification developed if necessary. The other problem, of
too many system states to test practically, is emerging in systems
engineering more generally as a significant issue (see for e.g.
\cite{50}). There is emerging a major research area concerned with
identifying new paradigms for verification.

\subsection{A sense of proportion}\label{a-sense-of-proportion}

On the one hand, the analysis above indicates significant problems to be
overcome in order to systems engineer QEST, on the other hand, various
in-depth analyses \cite{16,11} have forecast the delivery of viable
QEST within the next few years. The resolution of this paradox is that
systems engineering generally reduces cost, risk, and time in complex
projects. Whilst the delivery of viable QEST is possible, in common with
all disruptive technologies, the risks are comparatively high and
success is far from guaranteed. The near-term delivery of QEST will come
from early adopters. Thus quantum systems engineers need to consider the
techniques and processes that will enable the near-term delivery, whilst
developing the methods and applications that will be needed to reach the
initial maturity delivery of QEST in the medium term. The achievement of
this objective is not only technical; it will also require development
of the appropriate business models through which enterprises can assure
commercial success.

\section{Conclusions}\label{conclusions}

Systems engineering is an essential approach to the creation and
continued use of complex systems. Non-Classical Quantum Technologies
(QEST), which rely on manipulation of quantum states and exploitation of
quantum superposition and entanglement, introduce new complexities that
engineers have not had to deal with hitherto. However, the exciting new
capabilities that QEST may enable, and the advances of quantum mechanics
in recent years, suggest that we are on the brink of a quantum
revolution and must prepare to transition exciting quantum science into
viable engineered devices. Collaboration between quantum scientist and
systems engineers is needed to develop the engineering processes and
methods to realize the potential offered by QEST,

Using the Systems and Software Engineering - Systems Life Cycle
Processes standard \cite{5} as an analysis framework, we have analyzed
the extent to which systems engineering processes can meet this
challenge and identified a number of areas in which advances are
necessary. At a fundamental level, there is a need for a consistent and
reliable approach for defining the System of Interest (SOI) for QEST and
there is a significant lack of appropriate models to enable many of the
process stages. The important activity of system verification is
unsuitable for QEST using current approaches and methods; it is
suggested that an entirely new verification paradigm may be needed.

\acknowledgments

  Content includes material subject to \textcopyright\ Crown copyright (2017), Dstl.
  This material is licensed under the terms of the Open Government
  Licence except where otherwise stated.
\bibliographystyle{unsrtnat}
\bibliography{refs}

\begin{thebibliography}{69}
\providecommand{\natexlab}[1]{#1}
\providecommand{\url}[1]{\texttt{#1}}
\expandafter\ifx\csname urlstyle\endcsname\relax
  \providecommand{\doi}[1]{doi: #1}\else
  \providecommand{\doi}{doi: \begingroup \urlstyle{rm}\Url}\fi

\bibitem[Note1()]{Note1}
Note1.
\newblock It is appreciated that revolutions are usually defined
  retrospectively and that predicting them tends to be an unreasonably
  speculative endeavor.

\bibitem[Pritchard and Till(2014)]{1}
J.~Pritchard and S.~Till.
\newblock \emph{UK Quantum Technology Landscape 2014}.
\newblock Dstl, London, 2014.

\bibitem[Till and Pritchard(2016)]{2}
S.~Till and J.~Pritchard.
\newblock \emph{UK Quantum Technology Landscape 2016}.
\newblock Dstl, London, 2016.

\bibitem[Everitt et~al.(2016)Everitt, Henshaw, and Dwyer]{3}
M.~J. Everitt, M.~J. D.~C. Henshaw, and V.~M. Dwyer.
\newblock Quantum systems engineering: A structured approach to accelerating
  the development of a quantum technology industry.
\newblock In \emph{International Conference on Transparent Optical Networks}.
  IEEE, 2016.

\bibitem[Christensen(1997)]{4}
C.~M. Christensen.
\newblock \emph{The Innovator's Dilemma}.
\newblock Harvard Business School Press, 1997.

\bibitem[ISO/IEC/IEEE(2015)]{5}
ISO/IEC/IEEE.
\newblock Systems and software engineering --- system life cycle processes.
\newblock 15288, 2015.

\bibitem[Flood and Carson(1988)]{6}
R.~L. Flood and R.~E. Carson.
\newblock \emph{Dealing with complexity: an introduction to the theory and
  application of systems science}.
\newblock Plenum Press, New York, 1988.

\bibitem[Blanchard and Fabryscy(2013)]{7}
B.~S. Blanchard and W.~J. Fabryscy.
\newblock \emph{Systems Engineering and Analysis}.
\newblock Pearson Education, 5 edition, 2013.

\bibitem[Mankins(1995)]{8}
J.~C. Mankins.
\newblock Technology readiness levels.
\newblock 1995.

\bibitem[Note2()]{Note2}
Note2.
\newblock We are indebted to Andy Powell of Innovate UK KTN for this
  consideration.

\bibitem[Honour(2013)]{9}
E.~C. Honour.
\newblock \emph{Systems engineering return on investment}.
\newblock Univ. South Australia, 2013.

\bibitem[Allhoff et~al.(2010)Allhoff, Lin, and Moore]{10}
F.~Allhoff, P.~Lin, and D.~Moore.
\newblock \emph{What is Nanotechnology and why does it matter? From science to
  ethics}.
\newblock John Wiley \& Sons, New York, 2010.

\bibitem[Walport and Knight(2016)]{11}
S.~M. Walport and S.~P.~L. Knight.
\newblock \emph{The Quantum Age: technological opportunities}.
\newblock Government Office for Science, London, 2016.

\bibitem[Note3()]{Note3}
Note3.
\newblock This is somewhat of an oversimplifications but a discussion of some
  of the more subtle aspects of quantum physics are not within the scope of
  this paper.

\bibitem[O'Connell et~al.(2010)O'Connell, Hofheinz, Ansmann, Bialczak,
  Lenander, Lucero, Neeley, Sank, Wang, Weides, Wenner, Martinis, and
  Cleland]{12}
D.~O'Connell, M.~Hofheinz, M.~Ansmann, R.~C. Bialczak, M.~Lenander, E.~Lucero,
  M.~Neeley, D.~Sank, H.~Wang, M.~Weides, J.~Wenner, J.~M. Martinis, and
  N.~Cleland.
\newblock Quantum ground state and single-phonon control of a mechanical
  resonator.
\newblock \emph{Nature}, 464\penalty0 (7289):\penalty0 697--703, 2010.

\bibitem[Note4()]{Note4}
Note4.
\newblock Again a necessary simplification has been made.

\bibitem[Lee et~al.(2011)Lee, Sprague, Sussman, Nunn, Langford, m.~Jin,
  Champion, Michelberger, Reim, England, Jaksch, and Walmsley]{13}
K.~C. Lee, M.~R. Sprague, B.~J. Sussman, J.~Nunn, N.~K. Langford, X.~m.~Jin,
  T.~Champion, P.~Michelberger, K.~F. Reim, D.~England, D.~Jaksch, and I.~A.
  Walmsley.
\newblock Entangling macroscopic diamonds at room temperature.
\newblock \emph{Science}, 334\penalty0 (6060):\penalty0 1253--1256, 2011.

\bibitem[Peev et~al.(2009)Peev, Pacher, and All{\'e}aume]{14}
M.~Peev, C.~Pacher, and R.~All{\'e}aume.
\newblock The secoqc quantum key distribution network in vienna.
\newblock \emph{New J. Phys}, 11:\penalty0 75001, 2009.

\bibitem[Cayford et~al.(2015)Cayford, Gulijk, and Gelder]{15}
M.~Cayford, C.~Van Gulijk, and P.~H. A. J. M.~Van Gelder.
\newblock All swept up: An initial classification of nsa surveillance
  technology.
\newblock In \emph{Safety and Reliability: Methodology and Applications -
  Proceedings of the European Safety and Reliability Conference, ESREL 2014},
  pages 643--650, 2015.

\bibitem[Seffers(2016)]{16}
B.~G.~I. Seffers.
\newblock Quantum radar could render stealth aircraft obsolete.
\newblock \emph{SIGNAL}, 2016.

\bibitem[Tan et~al.(2008)Tan, Erkmen, Giovannetti, Guha, Lloyd, Maccone,
  Pirandola, and Shapiro]{16a}
Si-Hui Tan, Baris~I. Erkmen, Vittorio Giovannetti, Saikat Guha, Seth Lloyd,
  Lorenzo Maccone, Stefano Pirandola, and Jeffrey~H. Shapiro.
\newblock Quantum illumination with gaussian states.
\newblock \emph{Phys. Rev. Lett.}, 101:\penalty0 253601, Dec 2008.
\newblock \doi{10.1103/PhysRevLett.101.253601}.
\newblock URL \url{https://link.aps.org/doi/10.1103/PhysRevLett.101.253601}.

\bibitem[Note5()]{Note5}
Note5.
\newblock Sympathetic interactions and parasitic effects are the nomenclature
  terms used to broadly describe the undesired coupling of quantum states to
  unmitigated environmental interactions.

\bibitem[Bongs and Attallah(2017)]{17}
K.~Bongs and M.~Attallah.
\newblock Qt-shield: Compact lightweight high performance magnetic shielding
  enabling portable \& miniaturised quantum technology systems, 2017.
\newblock URL
  \url{{http://gow.epsrc.ac.uk/NGBOViewGrant.aspx?GrantRef=EP/R002789/1}}.

\bibitem[Note6()]{Note6}
Note6.
\newblock A group of technologies, or a major system, used as the base upon
  which applications are operated.

\bibitem[Fang and Qin(2012)]{18}
J.~Fang and J.~Qin.
\newblock Advances in atomic gyroscopes: A view from inertial navigation
  applications.
\newblock \emph{Sensors}, 12\penalty0 (5):\penalty0 6331--6346, 2012.

\bibitem[Parkinson and Spilker(1996)]{19}
B.~W. Parkinson and James~J. Spilker.
\newblock \emph{Global Positioning System: Theory and Applications, Volume 1}.
\newblock 1996.
\newblock ISBN 978-1-56347-106-3.

\bibitem[Woodman(2007)]{20}
O.~J. Woodman.
\newblock \emph{An introduction to inertial navigation}.
\newblock Cambridge University, Cambridge UK, 2007.

\bibitem[Jekeli(2005)]{21}
C.~Jekeli.
\newblock Navigation error analysis of atom interferometer inertial sensor.
\newblock \emph{Navigation}, 52\penalty0 (1):\penalty0 1--14, March 2005.

\bibitem[Battelier(2016)]{22}
B.~Battelier.
\newblock \emph{Development of compact cold-atom sensors for inertial
  navigation}.
\newblock SPIE Photonics Europe, 2016.

\bibitem[Barrett et~al.(2016)Barrett, Antoni-Micollier, Chichet, Battelier,
  L{\'e}v{\`e}que, Landragin, and Bouyer]{23}
B.~Barrett, L.~Antoni-Micollier, L.~Chichet, B.~Battelier, T.~L{\'e}v{\`e}que,
  A.~Landragin, and P.~Bouyer.
\newblock Dual matter-wave inertial sensors in weightlessness.
\newblock \emph{Nature Comms}, pages 1--11, 2016.

\bibitem[Bate(2013)]{24}
K.~Bate.
\newblock The silent service.
\newblock \emph{CIBSE J}, pages 22--27, Oct 2013.

\bibitem[Perez et~al.(2014)Perez, Salim, Farkas, Duggan, Ivory, and
  Anderson]{25}
Maximillian~A. Perez, Evan Salim, Daniel Farkas, Janet Duggan, Megan Ivory, and
  Dana Anderson.
\newblock On-chip optical trapping for atomic applications.
\newblock \emph{Proc.SPIE}, 9164:\penalty0 9164 -- 9164 -- 5, 2014.
\newblock \doi{10.1117/12.2064311}.
\newblock URL \url{http://dx.doi.org/10.1117/12.2064311}.

\bibitem[Keil et~al.(2016)Keil, Amit, Zhou, Groswasser, Japha, and Folman]{26}
M.~Keil, O.~Amit, S.~Zhou, D.~Groswasser, Y.~Japha, and R.~Folman.
\newblock Fifteen years of cold matter on the atom chip: Promise, realizations,
  and prospects.
\newblock \emph{J. Mod. Opt}, 340:\penalty0 1--49, July 2016.

\bibitem[Schkolnik et~al.(2016)Schkolnik, Hellmig, Wenzlawski, Grosse,
  Kohfeldt, D{\"o}ringshoff, Wicht, Windpassinger, Sengstock, Braxmaier,
  Krutzik, and Peters]{27}
V.~Schkolnik, O.~Hellmig, A.~Wenzlawski, J.~Grosse, A.~Kohfeldt,
  K.~D{\"o}ringshoff, A.~Wicht, P.~Windpassinger, K.~Sengstock, C.~Braxmaier,
  M.~Krutzik, and A.~Peters.
\newblock A compact and robust diode laser system for atom interferometry on a
  sounding rocket.
\newblock \emph{Appl. Phys. B Lasers Opt}, 122:\penalty0 8, 2016.

\bibitem[Papadakis(2012)]{28}
S.~J. Papadakis.
\newblock 5 scaling.
\newblock In M.~A.~Garrison Darrin and L.~Barth Janet, editors, \emph{Systems
  Engineering for Microscale and Nanoscale Technologies}. CRC Press p. 105,
  Boca Raton, 2012.

\bibitem[Jennifer(2012)]{29}
L.~Sample Jennifer.
\newblock 7 introduction to nanotechnology.
\newblock In M.~A.~Garrison Darrin and L.~Barth Janet, editors, \emph{Systems
  Engineering for Microscale and Nanoscale Technologies}. CRC Press, Boca
  Raton, 2012.

\bibitem[Breidenich(2012)]{30}
J.~Breidenich.
\newblock 11 interfaces at the micro- and nanoscale.
\newblock In M.~A.~Garrison Darrin and L.~Barth Janet, editors, \emph{Systems
  Engineering for Microscale and Nanoscale Technologies}. CRC Press, Boca
  Raton, 2012.

\bibitem[Kinder et~al.(2012)Kinder, Barot, Henshaw, and Siemieniuch]{31}
A.~Kinder, V.~Barot, M.~Henshaw, and C.~Siemieniuch.
\newblock System of systems:`defining the system of interest.
\newblock In \emph{Proc. of the 2012 7th International Conference on Systems of
  Systems Engineering}, pages 463--468. IEEE, 2012.

\bibitem[Boardman and Sauser()]{32}
J.~Boardman and B.~Sauser.
\newblock System of systems - the meaning of of.
\newblock In \emph{IEEE/SMC International Conference on System of Systems
  Engineering}.

\bibitem[Phillips(1998)]{33}
W.~D. Phillips.
\newblock Laser cooling and trapping of neutral atoms laser cooling.
\newblock \emph{Rev. Mod. Physic}, 70\penalty0 (3):\penalty0 1--20, 1998.

\bibitem[Flood(1987)]{34}
R.~L. Flood.
\newblock Some theoretical considerations of mathematical modeling.
\newblock \emph{in}, 31:\penalty0 354--360, 1987.

\bibitem[Darrin and Barth(2012)]{35}
Ann~Garrison Darrin and Janet~L. Barth.
\newblock Systems engineering for micro- and nanoscale technologies.
\newblock In \emph{Systems Engineering for Microscale and Nanoscale
  Technologies}. CRC Press, 2012.
\newblock ISBN 9781439837320.

\bibitem[Maranchi(2012)]{36}
J.~P. Maranchi.
\newblock Nanoscale systems - top-down assembly.
\newblock In \emph{Systems Engineering for Microscale and Nanoscale
  Technologies}. CRC Press, 2012.

\bibitem[Graedel et~al.(2015)Graedel, Harper, Nassar, Nuss, and Reck]{37}
T.~E. Graedel, E.~M. Harper, N.~T. Nassar, Philip Nuss, and Barbara~K. Reck.
\newblock Criticality of metals and metalloids.
\newblock \emph{Proceedings of the National Academy of Sciences}, 112\penalty0
  (14):\penalty0 4257--4262, 2015.
\newblock \doi{10.1073/pnas.1500415112}.
\newblock URL \url{http://www.pnas.org/content/112/14/4257.abstract}.

\bibitem[Burge(2007)]{38}
S.~Burge.
\newblock A functional approach to quality function deployment.
\newblock \emph{Syst. Eng}, pages 1--33, January 2007.
\newblock URL
  \url{https://www.burgehugheswalsh.co.uk/uploaded/1/documents/a-functional-approach-to-quality-function-deployement-v3.pdf}.

\bibitem[Cohen(1995)]{39}
L.~Cohen.
\newblock \emph{Quality Function Deployment - How to make FD work for you}.
\newblock Addison-Wesley Pub. Co, Reading, Mass, 1995.

\bibitem[Laboratory(2014)]{40}
National~Physical Laboratory.
\newblock Quantum timing, navigation and sensing showcase.
\newblock 2014.

\bibitem[Daw(2004)]{41}
A.~J. Daw.
\newblock New process and structure thinking for capability development.
\newblock In \emph{9th International Command and Control Research and
  Technology Symposium}, pages 1--18, 2004.
\newblock URL \url{http://www.dodccrp.org/events/9th_ICCRTS/CD/papers/027.pdf}.

\bibitem[Henshaw et~al.(2011)Henshaw, Baker, Carter, Colby, Cooper, Dickerson,
  Drawbridge, Evans, Fagg, Hart, Haywood-Evans, Hobbs, Huggett, Kinder, Morua,
  Pearce, Rabbets, Siemieniuch, Sinclair, Spencer, and Tutt]{42}
M.~Henshaw, S.~Baker, J.~Carter, S.~Colby, R.~Cooper, C.~Dickerson,
  N.~Drawbridge, C.~Evans, C.~Tech~John Fagg, S.~Hart, M.~Rachel Haywood-Evans,
  J.~Hobbs, S.~David Huggett, A.~Kinder, M.~Morua, D.~Pearce, T.~Rabbets,
  C.~Siemieniuch, K.~Sinclair, J.~Spencer, and S.~Tutt.
\newblock Assessment of open architectures within defence procurement.
\newblock 2011.
\newblock URL \url{https://dspace.lboro.ac.uk/dspace-jspui/handle/2134/8828}.

\bibitem[Lalli et~al.(1997)Lalli, Kastner, and Hartt]{43}
V.~R. Lalli, R.~E. Kastner, and H.~N. Hartt.
\newblock \emph{Training Manual for Elements of Interface Definition and
  Control}.
\newblock NASA, 1997.
\newblock URL \url{https://ntrs.nasa.gov/search.jsp?R=19970018043}.

\bibitem[Spero et~al.(2014)Spero, Avera, Valdez, and Goerger]{44}
E.~Spero, M.~P. Avera, P.~E. Valdez, and S.~R. Goerger.
\newblock Tradespace exploration for the engineering of resilient systems.
\newblock \emph{Procedia Comput. Sci}, 28:\penalty0 591--600, 2014.

\bibitem[Institute(2016{\natexlab{a}})]{45}
European Telecommunications~Standards Institute.
\newblock Quantum-safe cryptography (qsc); quantum-safe algorithmic framework,
  2016{\natexlab{a}}.

\bibitem[Institute(2016{\natexlab{b}})]{46}
European Telecommunications~Standards Institute.
\newblock Quantum key distribution (qkd); component characterization:
  characterizing optical components for qkd systems, 2016{\natexlab{b}}.

\bibitem[Note7()]{Note7}
Note7.
\newblock European Telecommunications Standards Institute.

\bibitem[Pritchard and Till(2015)]{47}
J.~Pritchard and S.~Till.
\newblock Quantum technologies in mod's future research programme.
\newblock Loughborough, 2015.

\bibitem[Note8()]{Note8}
Note8.
\newblock International Technology Roadmap for Semi-conductors.

\bibitem[Institute(2016{\natexlab{c}})]{48}
European Telecommunications~Standards Institute.
\newblock Quantum key distribution (qkd); component characterization:
  characterizing optical components for qkd systems, 2016{\natexlab{c}}.
\newblock vol. 1, pp. 1--136.

\bibitem[Sewell et~al.(2013)Sewell, Napolitano, Behbood, Colangelo, and
  Mitchell]{49}
R.~J. Sewell, M.~Napolitano, N.~Behbood, G.~Colangelo, and M.~W. Mitchell.
\newblock Certified quantum non-demolition measurement of a macroscopic
  material system.
\newblock \emph{Nat. Photonics}, 7:\penalty0 517--520, 2013.

\bibitem[Note9()]{Note9}
Note9.
\newblock \protect \url {http://cordis.europa.eu/project/rcn/193797_en.html}.

\bibitem[Hafner-zimmermann and Henshaw(2017)]{50}
S.~Hafner-zimmermann and M.~J. D.~C. Henshaw.
\newblock \emph{The future of trans-Atlantic collaboration in modelling and
  simulation of Cyber-Physical Systems A Strategic Research Agenda for
  Collaboration}.
\newblock Steinbeis-edition, Stuttgart, Germany, 2017.
\newblock ISBN 978-3-95663-112-2.

\bibitem[Giere(2004)]{51}
R.~N. Giere.
\newblock How models are used to represent reality.
\newblock \emph{Philos. Sci}, 71\penalty0 (5):\penalty0 742--752, 2004.

\bibitem[Lindblad(1976)]{52}
G.~Lindblad.
\newblock On the generators of quantum dynamical semigroups.
\newblock \emph{Commun. Math. Phys}, 48\penalty0 (2):\penalty0 119, 1976.

\bibitem[Duffus et~al.(2016)Duffus, Bjergstrom, Dwyer, Samson, Spiller,
  Zagoskin, Munro, Nemoto, and Everitt]{Duffus1}
S.~N.~A. Duffus, K.~N. Bjergstrom, V.~M. Dwyer, J.~H. Samson, T.~P. Spiller,
  A.~M. Zagoskin, W.~J. Munro, Kae Nemoto, and M.~J. Everitt.
\newblock Some implications of superconducting quantum interference to the
  application of master equations in engineering quantum technologies.
\newblock \emph{Phys. Rev. B}, 94:\penalty0 064518, Aug 2016.
\newblock \doi{10.1103/PhysRevB.94.064518}.
\newblock URL \url{https://link.aps.org/doi/10.1103/PhysRevB.94.064518}.

\bibitem[Duffus et~al.(2017)Duffus, Dwyer, and Everitt]{Duffus2}
S.~N.~A. Duffus, V.~M. Dwyer, and M.~J. Everitt.
\newblock Open quantum systems, effective hamiltonians, and device
  characterization.
\newblock \emph{Phys. Rev. B}, 96:\penalty0 134520, Oct 2017.
\newblock \doi{10.1103/PhysRevB.96.134520}.
\newblock URL \url{https://link.aps.org/doi/10.1103/PhysRevB.96.134520}.

\bibitem[Bowen et~al.(2017)Bowen, Dwyer, Phillips, and Everitt]{53}
J.~J. Bowen, V.~M. Dwyer, I.~W. Phillips, and M.~J. Everitt.
\newblock On calculating the dynamics of very large quantum systems, 2017.
\newblock URL \url{https://arxiv.org/abs/1702.01723}.

\bibitem[Gough(2017)]{54}
John Gough.
\newblock The tyranny of qubits - quantum technology's scalability bottleneck,
  2017.
\newblock URL \url{https://arxiv.org/abs/1703.05342}.

\bibitem[Di{\'o}si(2014)]{55}
L.~Di{\'o}si.
\newblock Hybrid quantum-classical master equations.
\newblock \emph{Physica Scripta}, 2014\penalty0 (T163):\penalty0 014004, 2014.
\newblock URL \url{http://stacks.iop.org/1402-4896/2014/i=T163/a=014004}.

\bibitem[Sait and Yousef(1999)]{56}
S.~Sait and H.~Yousef.
\newblock \emph{VLSI Physical Design Automation: Theory and Practice}.
\newblock World Scientific, 1999.

\bibitem[Brook(2016)]{57}
P.~Brook.
\newblock On the nature of systems of systems.
\newblock \emph{INCOSE Ann. Symp}, 2016.
\newblock \doi{10.1002/j.2334-5837.2016.00240.x}.

\end{thebibliography}
\end{document}